\documentclass[twocolumn,aps,prc,showpacs,superscriptaddress,preprintnumbers,floatfix,nofootinbib]{revtex4}
\usepackage{epsfig,graphics}
\usepackage{graphicx}
\usepackage{dcolumn}
\usepackage{bm}
\usepackage{amsmath}
\usepackage[usenames]{color}
\usepackage{ulem} 

\begin{document}

\title{Correlation between elliptic flow and shear viscosity in intermediate-energy heavy-ion collisions}
\author{C. L. Zhou}
\affiliation{Shanghai Institute of Applied Physics, Chinese
Academy of Sciences, Shanghai 201800, China}
\author{Y. G.  Ma\footnote{Correspondening author.
ygma@sinap.ac.cn}}
\affiliation{Shanghai Institute of Applied Physics, Chinese
Academy of Sciences, Shanghai 201800, China}
\affiliation{Shanghai Tech University, Shanghai 200031, China}
\author{D. Q. Fang}
\affiliation{Shanghai Institute of Applied Physics, Chinese
Academy of Sciences, Shanghai 201800, China}
\author{G. Q. Zhang}
\affiliation{Shanghai Institute of Applied Physics, Chinese
Academy of Sciences, Shanghai 201800, China}
\author{J. Xu}
\affiliation{Shanghai Institute of Applied Physics, Chinese
Academy of Sciences, Shanghai 201800, China}
\author{X. G. Cao}
\affiliation{Shanghai Institute of Applied Physics, Chinese
Academy of Sciences, Shanghai 201800, China}
\author{W. Q.  Shen}
\affiliation{Shanghai Institute of Applied Physics, Chinese
Academy of Sciences, Shanghai 201800, China}
\affiliation{Shanghai Tech University, Shanghai 200031, China}

\date{\today}

\begin{abstract}
The correlation between the elliptic flow $v_2$ scaled by the impact
parameter $b$ and the shear viscosity $\eta$ as well as the specific
viscosity $\eta/s$, defined as the ratio of the shear viscosity to
the entropy density $s$, is investigated for the first time in
intermediate-energy heavy-ion collisions based on an
isospin-dependent quantum molecular dynamic model. The elliptic flow
is calculated at balance energies to exclude the geometric influence
such as the blocking effect from the spectators. Our study shows
that $v_{2}/b$ decreases almost linearly with increasing $\eta$,
consistent with that observed in ultra-relativistic heavy-ion
collisions. On the other hand, $v_{2}/b$ is found to increase with
increasing $\eta/s$.
\end{abstract}
\maketitle


One of the most important spot in heavy-ion collisions is the phase
transition of the strong interacting matter, e.g., the transition
between the quark-gluon plasma (QGP) and the hadronic matter at
ultra-relativistic energies as well as the liquid-gas phase
transition (LGP) at intermediate
energies~\cite{MaPRL,Moretto,STAR_PHENIX2005,Lacey2007,Bonasera2000,Borderie2008,Liu-Ko,zhoucl2013,zhoucl2012,lisx2011,Fang2014,XuPLB,XuNST}.
Recently, one interesting probe of the phase transition is the
so-called specific viscosity $\eta/s$, defined as the ratio of the
shear viscosity $\eta$ to the entropy density $s$. Empirical
observations of the temperature or incident energy dependence of
$\eta/s$ for H$_{2}$O, He, and Ne$_{2}$ exhibit a minimum in the
vicinity of the phase transition temperature~\cite{Csernai2006}.
Besides, with increasing temperature, it is found that $\eta/s$
decreases steeply below and rises slowly above the critical
temperature for a wide class of systems. Nevertheless, a lower bound
of $\eta/s \geq 1/4\pi$, obtained by Kovtun-Son-Starinets (KSS) for
infinitely coupled super-symmetric Yang-Mills gauge theory based on
the AdS/CFT duality conjecture, is speculated to be universally
valid~\cite{Kovtun2005}. It is thus very important to
study $\eta/s$ of the strong interacting matter created in heavy-ion
collisions at both ultra-relativistic and intermediate energies.

The most common approach to study transport properties of the QGP
formed in ultra-relativistic heavy-ion collisions is to investigate
the effects of $\eta/s$ on the elliptic flow in a viscous
hydrodynamic model~\cite{romatschke2007,song2008,huovinen2009}. This
approach has been implemented to simulate heavy-ion collisions at
Relativistic Heavy-Ion Collider (RHIC) and Large Hadron Collider
(LHC), and works well in mid-central collisions or at higher
energies. In peripheral collisions or at lower energies, however,
the hydrodynamic model should be followed by a microscopic transport
simulation for the highly dissipative hadronic
phase~\cite{song2011,ryu2012}. In addition, the hydrodynamics failed
to describe the intermediate-energy heavy-ion collisions that evolve
mainly in hadronic degrees of freedom. Therefore, no efforts have
been devoted to study the viscous effect on the elliptic flow in
intermediate-energy heavy-ion collisions. However, some transport
models ~\cite{zhoucl2013,zhoucl2012,lisx2011,Fang2014}, e.g., the isospin-dependent quantum molecular dynamic (IQMD)
model and the Boltzmann-Uehling-Uhlenbeck (BUU) model,  as well as an isospin- and
momentum-dependent interaction \cite{XuPLB,XuNST} can give very
reasonable results for intermediate-energy heavy-ion collisions. In
our previous studies, efforts
have been made to study the LGP through $\eta/s$ based on the IQMD
and BUU model, and a good agreement with previous analyses has been
found.

It is worth mentioning that the dynamic evolution is more complex in
low and intermediate-energy heavy-ion collisions. For instance, the
elliptic flow at low energies is related to the rotation of the
compound system with the expansion of the hot and compressed
participant zone and possibly modified by the shadowing effect of
the cold spectator matter. In heavy-ion collisions at
ultra-relativistic energies, however, the spectators are quickly
separated from the participants, and the participant part can evolve
without the influence of spectator matter. It is thus seen that the
elliptic flow is positive at low energies, becomes negative at
intermediate energies, and is positive again at relativistic
energies~\cite{Dan2002,FOPI,WangJia}. In transport model simulations at
intermediate-energy heavy-ion collisions, nucleon interactions are
described by mean fields as well as nucleon-nucleon (N-N)
scatterings in the whole reaction process. With the increasing
collision energy, the N-N scatterings become dominating for the
dynamics and are responsible for the deflection of the hot
compressed participant matter from the cold spectator matter, while
the attractive part of the mean field becomes more and more
important with decreasing collision energy. These observations
indicate that the collective flow of the participant part mainly
suffers the compression and pulling out effects from the spectator
matter. Therefore, the viscous effect can only be correctly studied
once the influence of the spectator on the elliptic flow has been
successfully removed.

Fortunately, the two competing effects of mean fields and N-N
scatterings largely balance each other at the so-called balance
energy, characterized by a vanishing direct
flow~\cite{bertsch1987,magestro2000,andronic2001,Puri2011}, i.e., when the
slope parameter $F_{d}$ of the directed flow denoting the average
transverse momentum in the reaction plane at mid rapidity becomes
zero. The influence of the spectator matter on the participant
matter is relatively small in collisions at the balance energy, and
these collisions provide an excellent opportunity to study the
viscous effect on the elliptic flow. One should keep in mind that
the referred balance between mean fields and N-N collisions does not
mean that they are balanced at the whole collision process. At the
maximum compressed stage, the N-N scatterings are stronger while the
mean fields take over after some time of expansion. This means that
the participant part is still influenced by the spectator matter in
the expansion stage. It is thus difficult to calculate reasonably
the conventional spatial-eccentricity-scaled elliptic flow
$v_{2}/\epsilon$~\cite{ferini2009,voloshin2000}, because the spatial
eccentricity, which is defined as $\epsilon=\langle y^2-x^2 \rangle
/\langle y^2+x^2 \rangle$ characterizing the anisotropy in the
coordinate space, changes through the whole collision process.
Different from that in ultra-relativistic heavy-ion collisions, an
impact-parameter-scaled elliptic flow $v_{2}/b$ is introduced to
eliminate the geometric effect in the present study.

Our study is based on an isospin-dependent quantum molecular dynamic
model (IQMD)~\cite{Hartnack1989,Hartnack1998,Aichelin1991}. In this
framework, each nucleon is represented by a Gaussian wave packet in
coordinate and momentum space to partially take into account the
quantum effects, and the equation of motion for the center of the
wave packet evolves according to the classical equation of motion
based on the Hamiltonian of the system from effective nucleon
interactions. The effective mean field used in the IQMD model can be expressed
as~\cite{Aichelin1991}
\begin{equation}
U = U_{\rm Sky} + U_{\rm Coul}  + U_{\rm Yuk} + U_{\rm sym},
\end{equation}
where $U_{\rm Sky}$, $U_{\rm Coul}$, $U_{\rm Yuk}$, and $U_{\rm sym}$ are the bulk Skyrme potential, the Coulomb potential, the
surface Yukawa potential, and the isospin asymmetry potential, respectively.
The bulk Skyrme potential is
\begin{equation}
U_{\rm Sky} = \alpha(\rho/\rho_{0}) + \beta{(\rho/\rho_{0})}^{\gamma},
\end{equation}
where $\rho$ is the nucleon number density and $\rho_{0}$ = 0.16 fm$^{-3}$ is the saturation density. In this work, the parameters $\alpha =-356$ MeV, $\beta =303$ MeV, and $\gamma = 7/6$, corresponding to a soft equation of state, are used. The expressions of the other potentials can be found in our previous work~\cite{zhoucl2013,zhoucl2012}. Within the present framework, reasonable phase-space information of nucleons and fragments in intermediate-energy heavy-ion collisions can be obtained. About 50,000 $^{197}$Au+$^{197}$Au collision events have been simulated using the IQMD model for different impact parameters and beam energies for the investigation of the correlation between the elliptic flow and the shear viscosity in the present study.

\begin{figure}
\includegraphics[width=8cm]{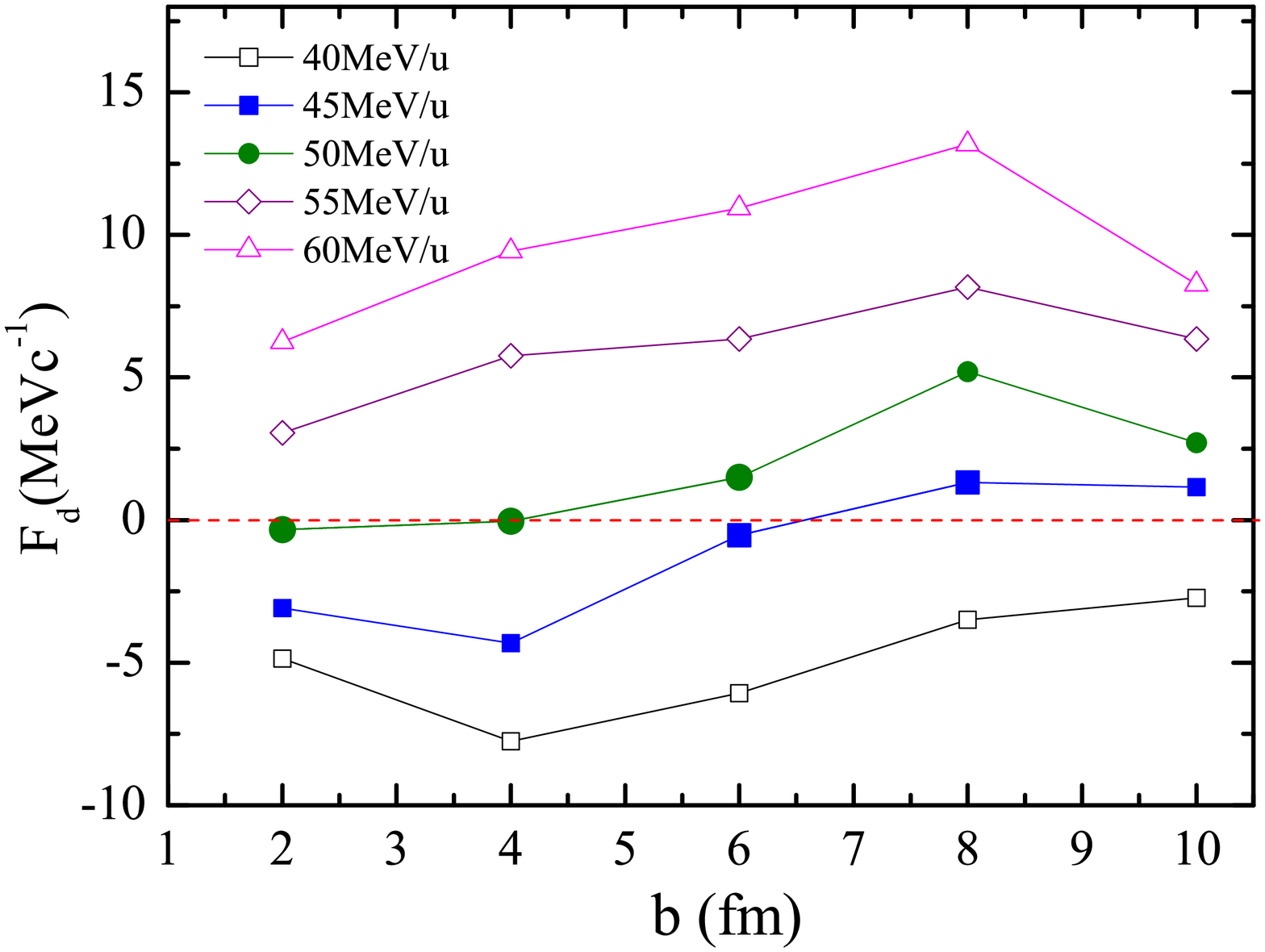}
\vspace{-0.1truein} \caption{\footnotesize (Color online) The
extracted slope parameter $F_d$ of the directed flow as a function of
the impact parameter $b$ in heavy-ion collisions at different beam
energies. }\label{Fig_v1-b}
\end{figure}

The slope parameter of the directed flow in non-central heavy-ion
collisions can be expressed as
\begin{equation}
F_{d}=\frac{d\langle p_{x}/A \rangle}{d(y/y_{b})},
\end{equation}
where $p_x$ is the projection of the transverse momentum in the
reaction plane, $A$ is the number of nucleons, and $y/y_{b}$ is the
particle rapidity $y$ normalized by the beam rapidity $y_{b}$. The
extracted $F_{d}$ from the IQMD model versus the impact parameter
$b$ at different beam energies is shown in Fig.~\ref{Fig_v1-b}. The
dashed line of $F_{d}=0$ is plotted to guide eyes. With the increase
of the impact parameter, it is seen that the $|F_{d}|$ generally
increases, passes through a maximum value, and diminishes in most
peripheral collisions~\cite{pak1996,soff1995}. Besides, $F_{d}$ is
negative at lower energy of $40$ MeV/u, positive at higher energies
of $55-60$ MeV/u, and zero around $45-50$ MeV/u, consistent with the
findings in Refs.~\cite{sood2004,crochet1997,zhang1990,partlan1995}.
The collision energies and impact parameters with $F_d$ most close
to $0$ are: [50 MeV/u, 2 fm], [50 MeV/u, 4 fm], [50 MeV/u, 6 fm],
[45 MeV/u, 6 fm], and [45 MeV/u, 8 fm], respectively. In these
collisions, the effect from the mean field and N-N scatterings
cancel each other so that the influence of the blocking from the
spectator matter on the elliptic flow are negligible.

\begin{figure}
\includegraphics[width=8cm,bb = 0 0 780 530]{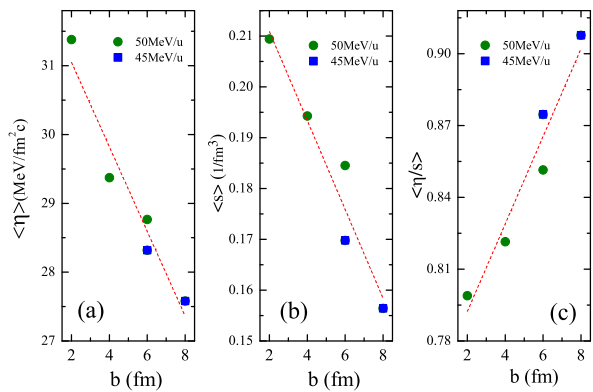}
\vspace{-0.1truein} \caption{\footnotesize (Color online) The
extracted average shear viscosity (a), entropy density (b), and
specific viscosity (c) at balance energies and corresponding impact
parameters. The dashed lines are plotted to guide eyes.
}\label{Fig_ets-b}
\end{figure}

For largely equilibrated systems, fluxes of macroscopic quantities
are proportional to the field gradient of the system. The shear
viscosity $\eta$ is the coefficient of proportionality between the
shear force between flow layers per unit area, i.e., the momentum
flux, and the velocity gradient, and it can be understood from the
fluctuation-dissipation theorem~\cite{kubo1996}. In the limit of
Boltzmann statistics, the shear viscosity corresponds to the
first-order Chapman-Enskog coefficient, and can be expressed as a
parameterized function by
Danielewicz~\cite{Danielewicz1984,Danielewicz2009}
\begin{eqnarray}
\eta\left(\frac{\rho}{\rho_{0}},T\right) &=& \frac{1700}{T^2}\left(\frac{\rho}{\rho_{0}}\right)^{2}+\frac{22}{1+T^{2}10^{-3}}\left(\frac{\rho}{\rho_{0}}\right)^{0.7}  \nonumber \\
&& + \frac{5.8\sqrt{T}}{1+160T^{-2}},
\label{Dani-eqn}
\end{eqnarray}
where the shear viscosity $\eta$ and the temperature $T$ are in
MeV/fm$^2 c$ and MeV, respectively. The above equation is reliable
if the system is locally equilibrated with density $\rho$ and
temperature $T$, with the former calculated from the overlap of the
nucleon wave packets, and the latter as well as the entropy density
$s$ obtained from the generalized Thomas-Fermi formulism at finite
temperature~\cite{khoa1992a,puri1992,barranco1981,rashdan1987}.
The average shear viscosity $\langle \eta \rangle$, entropy density
$\langle s \rangle $, and specific viscosity $\langle \eta /s
\rangle$ in the center of the system from the moment of maximum
compression to freeze-out for the five selected beam energies and
impact parameters are given in panel (a), (b), and (c) of
Fig.~\ref{Fig_ets-b}, respectively. It is interesting to see that
the $\langle \eta \rangle$ decreases with increasing impact
parameter. This can be understood from the classical relation $\eta
\sim \langle p \rangle/\sigma$, where the average momentum $\langle
p \rangle$ is larger in more central collisions while the N-N
scattering cross section $\sigma$ is the same. In addition, the
$\langle s \rangle $ is also larger in more central collisions due
to the higher temperature. After taking the ratio, the average
specific viscosity $\langle \eta  / s \rangle$ somehow increases
with increasing impact parameter.

\begin{figure}
\includegraphics[width=8cm]{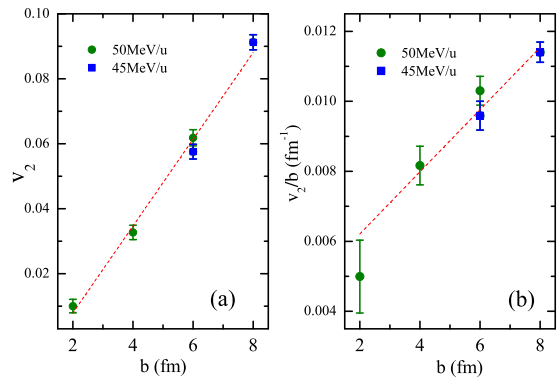}
\vspace{-0.1truein} \caption{\footnotesize (Color online) The
elliptic flow and that scaled by the impact parameter at balance
energies and corresponding impact parameters. The dashed lines are plotted to guide eyes.}\label{Fig_v2-b}
\end{figure}

The elliptic flow is defined as the second-order harmonic
coefficient of Fourier expansion of the particle azimuthal
distribution
\begin{equation}
v_{2}=\langle \cos(2\phi) \rangle = \left\langle \frac{p_{x}^2-p_{y}^2}{p_{x}^2+p_{y}^2} \right\rangle,
\end{equation}
where $\phi$ is the azimuthal angle, $p_{x}$ and $p_{y}$ are the
projections of the transverse momentum parallel and perpendicular to
the reaction plane, respectively, and the bracket denotes the
average over all the particles. The elliptic flow can be determined
by the collective motion resulting from the rotation of the compound
system, the expansion of the hot and compressed participant matter,
and the possible modification by the shadowing effect of the cold
spectator matter~\cite{wilson1990,tsang1993,lacey1993,wilson1995,shen1998,yumingzheng1999}.
Similar to the directed flow, generally the elliptic flow first
increases with increasing impact parameter, reaches a maximum in
mid-central collisions, and then decreases at large centralities. In
this work the emitted light fragments of charge number $Z\le3$
including protons and neutrons at mid-rapidity $|y/y_{b}|\le0.1$ are
employed to calculate the elliptic flow at freeze-out. The impact
parameter dependence of $v_{2}$ and that scaled by the impact
parameter is shown in panel (a) and panel (b) of
Fig.~\ref{Fig_v2-b}, respectively. The positive value of $v_{2}$
indicates that an in-plane emission of particles is observed at
balance energies. It is shown in panel (a) that $v_{2}$ increases
linearly with increasing impact parameter, and the $v_2$ at [45
MeV/u, 8 fm] is about 10 times that at [50 MeV/u, 2 fm]. After
scaled by the impact parameter, this difference is reduced to about
2 times as shown in panel (b), and we argue that the remaining
difference is due to the viscous effect in intermediate-energy
heavy-ion collisions.

\begin{figure}
\includegraphics[width=8cm]{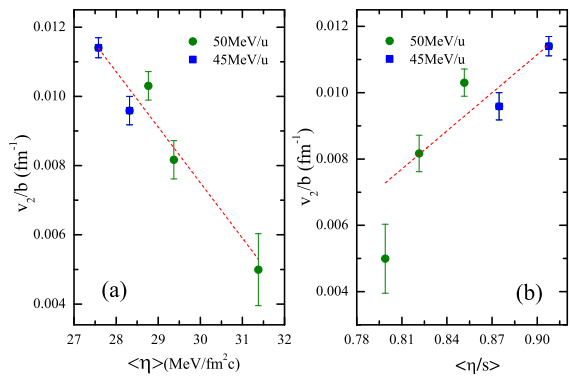}
\vspace{-0.1truein} \caption{\footnotesize (Color online) The
scaled elliptic flow as a function of average shear viscosity and specific viscosity
at balance energies and corresponding impact parameters. The dashed lines are plotted to guide eyes. }\label{Fig_v2/b-ets}
\end{figure}

Combing Fig.~\ref{Fig_ets-b} and Fig.~\ref{Fig_v2-b}, the
correlations between $v_{2}/b$ and $\langle \eta \rangle$ as well as
$\langle \eta / s \rangle$ are exhibited in panel (a) and (b) of
Fig.~\ref{Fig_v2/b-ets}, respectively. It is found that $v_{2}/b$
decreases almost linearly with increasing average shear viscosity
$\langle \eta \rangle$. This shows that a stronger interaction,
which leads to a smaller shear viscosity, is more efficient in
transforming the initial eccentricity to the final elliptic flow,
consistent with the findings in heavy-ion collisions at
ultra-relativistic energies. On the other hand, $v_{2}/b$ somehow
mostly increases with increasing average specific viscosity $\langle
\eta / s \rangle$, different from that in ultra-relativistic
heavy-ion collisions. This might be due to the stronger dissipation
in the hadronic phase than in the partonic phase, which leads to a
different behavior of the entropy density in the former case.

In summary, the correlations between the impact-parameter-scaled
elliptic flow $v_{2}/b$ and the shear viscosity $\eta$ as well as
the specific viscosity $\eta/s$ are investigated based on an
isospin-dependent quantum molecular dynamic model. Specific
combinations of energy and impact parameter, at which the direct
flow disappears, are selected in $^{197}$Au+$^{197}$Au collisions to
remove the blocking effects of the cold spectator matter on the
elliptic flow. The shear viscosity is calculated from the
parameterized formulism by Danielewicz for the participant nuclear
matter, and the local density, temperature, and entropy density are
extracted from the hot Thomas-Fermi formulism. Our calculation shows
the scaled elliptic flow from the light fragments with charge number
$Z\le3$ decreases within increasing shear viscosity, consistent with
that observed in ultra-relativistic heavy-ion collisions at RHIC or
LHC. On the other hand, $v_{2}/b$ increases with increasing specific
viscosity. Our findings are useful in extracting experimentally the
shear viscosity and specific viscosity from the elliptic flow in
heavy-ion collisions at balance energies.

This work is partially supported by the NSFC under contracts No.11035009, 11220101005, 10979074, 11175231, 11405248,
the Major State Basic Research Development Program in China under Contract No. 2014CB845401, 2013CB834405, the "100-talent plan" of Shanghai Institute of Applied Physics under grant Y290061011 from the Chinese Academy of Sciences, and the Knowledge Innovation Program of Chinese Academy of Science.


\end{document}